# A compact, low field, broadband matching section for externally-powered X-band dielectric-loaded accelerating structures

Yelong Wei, [1,*] Alexej Grudiev [1], Ben Freemire [2], and Chunguang Jing [2]

[1]CERN, Geneva CH-1211, Switzerland

[2]Euclid Techlabs LLC, Bolingbrook, Illinois 60440, USA

Abstract: It has been technically challenging to efficiently couple external radiofrequency (RF) power to cylindrical dielectric-loaded accelerating (DLA) structures, especially when the DLA structure has a high dielectric constant. This paper presents a novel design of matching section for coupling the RF power from a circular waveguide to an X-band DLA structure with a dielectric constant $\varepsilon_r = 16.66$ and a loss tangent $\tan\delta = 3.43 \times 10^{-5}$. It consists of a very compact dielectric disk with a width of 2.035 mm and a tilt angle of $60^0$, resulting in a broadband coupling at a low RF field which has the potential to survive in the high-power environment. To prevent a sharp dielectric corner break, a $45^0$ chamfer is also added. A microscale vacuum gap, caused by metallic clamping between the thin coating and the outer thick copper jacket, is also studied in detail. Through optimizations, more than 99% of RF power is coupled into the DLA structure, with the maximum electromagnetic fields located at a DLA structure. Tolerance studies on the geometrical parameters and mechanical design of the full-assembly structure are also carried out as a reference for realistic fabrication.

## I. Introduction

Dielectric-loaded accelerating (DLA) structures, utilizing dielectrics to slow down the phase velocity of travelling waves in the vacuum channel, have been studied both theoretically [1-5] and experimentally [6-10] as a potential alternative to conventional disk-loaded copper structures. A DLA structure has a simple geometry comprising a dielectric tube surrounded by a conducting cylinder. The simplicity of DLA structures offers great advantages for fabrication of high frequency (>10 GHz) accelerating structures, as compared with conventional metallic structures which demand extremely tight fabrication tolerances. This is of great importance in the case of linear colliders, where tens of thousands of accelerating structures have to be built. Moreover, the relatively small diameter of DLA structures facilitates the use of quadrupole lenses around the structures. The DLAs are also advantageous in terms of the ease of applying damping schemes for beam-induced deflection modes [11-12], which can cause bunch-to-bunch beam breakup and intrabunch head-tail instabilities [13]. However, there are still some potential challenges for DLA structures in a high-power RF environment, such as breakdown, thermal heating and multipactor [8, 14-18]. In dielectric breakdown studies, a dielectric surface field breakdown threshold of 13.8 GV/m has been observed [19]. Although there are no direct breakdown studies for externally powered DLA structures, no breakdown has been observed in several high power tests carried out on a DLA structure at a level > 5 MW [14-15, 17-18]. Thermal heating issues can be solved by achieving good contact between the ceramics coating and the outer copper jacket connected to the cooling system. The surface resonant multipactor has been always observed in experimental studies to absorb a large fraction of the incident RF power, which is identified as an issue limiting the gradient in DLA structures. An effective approach which uses an applied axial magnetic field to completely suppress this multipactor in DLA structures has been proposed [16] and demonstrated [17-18] in high-power experimental studies.

*yelong.wei@cern.ch

In addition to these potential challenges, a practical issue to be addressed is the efficient coupling of the RF power into a DLA structure with an outer diameter much smaller than the rectangular waveguide. A scheme [20] using a combination of a side coupling slot and a tapered dielectric layer near the slot was proposed to couple the RF power from a rectangular waveguide into the DLA structure. The tapered dielectric matching section is a part of the DLA structure. It converts $TE_{10}$ mode from a rectangular waveguide to $TM_{01}$ mode in the circular dielectric-loaded waveguide. A power coupling coefficient (it is defined by $\eta = 10^{-\frac{S_{21}}{10}}$ and frequently used in the analysis in this paper) greater than 95% can be achieved by carefully tuning the coupling slot and tapering the inner radius of the dielectric tube near the coupling slot. However, breakdown was observed for such a RF coupler in the high-power test because of the strong electric field enhancement near the slot [21]. In order to eliminate any field enhancement near the RF coupler due to the presence of dielectrics, another coupling scheme was adopted to separate the RF coupler from the DLA structure [22-23]. There are 3 modules in this scheme: a RF coupler section, a tapered dielectric matching section, and a DLA section. The RF coupling section is used to convert the $TE_{10}$ mode from a rectangular waveguide to the $TM_{01}$ mode in a circular waveguide. The tapered dielectric section provides a good match for the impedance of the $TM_{01}$ mode between the circular waveguide and the dielectric-loaded waveguide. This scheme separates the dielectric-loaded waveguide from the RF coupler by a tapered matching section and thus makes the RF coupler independent of the dielectric properties. In this case, the RF coupler, which works as a mode-converter, can be reused for similar experiments operating at the same frequency. The area of the coupling slot is also much larger, enabling a RF field in the coupling slot much lower than that in the former scheme, under the same input power. Such a tapered dielectric matching section has been used for many high-power experimental studies [8, 14-15, 17-18]. In [8, 17] strong multipactor was observed in a tapered dielectric matching section with a length of >30 mm. At this length, it is a big challenge to completely suppress multipactor, requiring a large solenoidal magnet capable of producing a uniform axial magnetic field over the whole DLA structure with a tapered dielectric matching section. This length also occupies valuable space which could be saved for accelerating structures. Thus, a compact matching section with a much shorter length would, if realized, represent an advance for externally powered DLA structures.

Motivated by these points, we present in this paper a novel design of matching section for efficiently coupling the RF power from a circular waveguide to an X-band DLA structure. There are also 3 modules in our RF coupling scheme, as shown in Fig. 1. The mode converter has been studied in detail [24], so we concentrate our efforts on the design of the matching section and DLA structure. Section II presents detailed studies for a DLA structure with a dielectric constant $\varepsilon_r = 16.66$ and a loss tangent $\tan\delta = 3.43 \times 10^{-5}$. Section III describes the RF design for a dielectric matching section to achieve the best coupling with the maximum fields located at the DLA structure. Section IV investigates the chamfer of sharp dielectric corner and a microscale vacuum gap caused by metallic clamping from the point of view of realistic fabrication. Section V shows the tolerance studies for the geometrical parameters of the matching section. Section VI gives the RF performance for the mechanical full-assembly structure, including the DLA structure connected with two matching sections, circular waveguides with the choke geometry, and the $TE_{10}$-$TM_{01}$ mode converters.

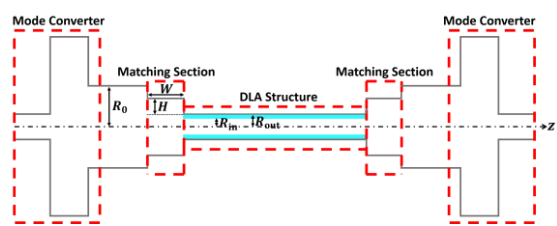

Figure 1. Conceptual illustration of an externally-powered DLA structure, two $TE_{10}$-$TM_{01}$ mode converters and two matching sections. $R_0$, $W$, and $H$ represent circular waveguide radius, matching section width, and matching section height, respectively, while $R_{in}$ and $R_{out}$ represent inner radius and outer radius for the DLA structure.

## II. Design of a DLA structure

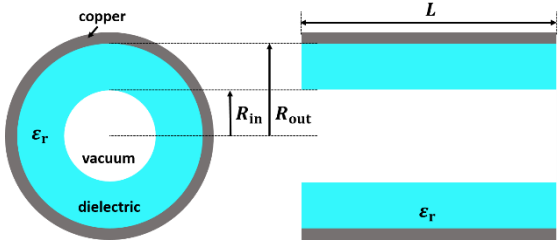

Figure 2. Front view and longitudinal cross section of a cylindrical DLA structure. $\varepsilon_r$, $R_{in}$, $R_{out}$, and $L$ represent dielectric constant, inner radius, outer radius, and length for the DLA structure.

In this section, the RF properties of a DLA structure are studied in detail. A DLA structure comprises a uniform linear dielectric tube surrounded by a copper cylinder, as shown in Fig. 2. Unlike conventional metallic accelerating structures, the uniform DLA structure does not have any geometrical periodicity. By adjusting dielectric constant $\varepsilon_r$, inner radius $R_{in}$, and outer radius $R_{out}$, the DLA structure can be operated as a slow-wave constant-impedance accelerator. The ceramic materials for dielectric-based accelerating structures have to withstand high accelerating fields, prevent potential charging by particle beams, have good thermal conductivity, and generate low power loss. $MgTiO_3$ ceramic, with a dielectric constant $\sim 16$ and an ultralow loss tangent $\tan\delta \leq 1.0 \times 10^{-4}$, which has been studied in [25-26], is chosen as the dielectric material for such a DLA structure.

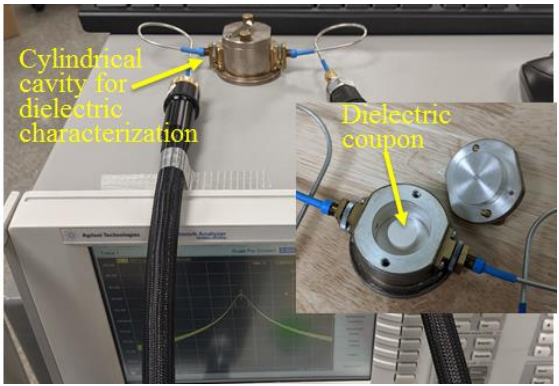

Figure 3. Measurement setup of the dielectric properties of the material sample.

An accurate measurement of the dielectric properties has to be performed before using such a material for our RF design. As shown in Fig. 3, a $TE_{01\delta}$ silver-plated resonator with a high quality factor, which is designed for testing ceramics at an X-band frequency, is used to measure the dielectric constant $\varepsilon_r$ and loss tangent $\tan\delta$ of sample coupons. Four dielectric coupons made from the same dielectric rods as for the fabrication of the DLA structure are measured. A dielectric constant $\varepsilon_r = 16.66$ and an ultralow loss tangent $\tan\delta = 3.43 \times 10^{-5}$ (having error bars 0.6% of the nominal value) are obtained for the RF design of the DLA structure and matching sections which follows.

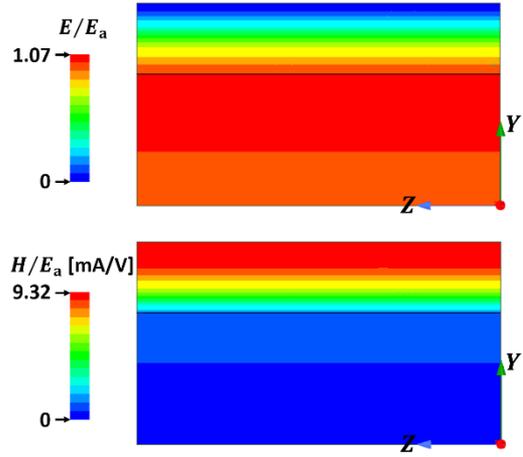

Figure 4. (a) Electric field distribution $E/E_a$, and (b) magnetic field distribution $H/E_a$ for the accelerating $TM_{01}$ mode in a DLA structure.

HFSS [27] is used to compute the electromagnetic fields for this DLA structure. The ratio of the peak electric field $E_p$ to the average accelerating field $E_a$ limits the achievable accelerating gradient for conventional iris-loaded metallic structures. Typically this ratio $E_p/E_a \geq 2$ [28-29]. Figure 4 shows the electric field distribution $E/E_a$ and magnetic field distribution $H/E_a$ of the $TM_{01}$ mode in a DLA structure, where $E$, $E_a$, $H$ represent the electric field, the average accelerating field, and the magnetic field, respectively. As shown in Fig. 4(a), the ratio of the peak electric field to the average accelerating field for our DLA structure is calculated to be 1.07, indicating that the accelerating gradient achieved is potentially higher than that of conventional metallic structures. The strongest electric field of $1.07E_a$ is located in the vacuum region next to the dielectric surface while the axial electric field is $E_a$. This slight difference along the *y* axis allows the accelerating fields in the vacuum region to be almost identical and uniform. Figure 4(b) shows that for the same DLA structure, the ratio of the peak magnetic field to the average accelerating field is 9.32 mA/V, which is much larger than that of existing metallic CLIC-G [30-32] structures. This means that strong magnetic fields are located on the metallic surface, resulting in large surface currents and hence high power loss and high pulse surface heating temperature rise [33-35].

Table 1. Optimum parameters for a DLA structure operating at $TM_{01}$ mode.

| Parameters | A DLA structure |
| --- | --- |
| Dielectric constant $\varepsilon_r$ | 16.66 |
| Dielectric loss tangent $\tan \delta$ | $3.43 \times 10^{-5}$ |
| Inner radius $R_{\text{in}}$ [mm] | 3.0 |
| Outer radius $R_{\text{out}}$ [mm] | 4.6388 |
| Length $L$ [mm] | 100.0 |
| Acceleration mode | $TM_{01}$ |
| Frequency $f$ [GHz] | 11.994 |
| Unloaded quality factor $Q_0$ | 2829 |
| Shunt impedance $r'$ [MΩ/m] | 26.5 |
| Group velocity $v_g/c$ | 0.066 |
| $E_s/E_a$ | 1.07 |
| $H_s/E_a$ [mA/V] | 9.32 |
| Power required to generate a gradient of 100 MV/m [MW] | 280 |

Table 1 shows all the geometrical and RF parameters for our DLA structure operating in $TM_{01}$ mode. The inner radius is chosen to be 3.0 mm from consideration of the beam dynamics requirement of CLIC designs [30-32]. The outer radius is then calculated to be 4.6388 mm for an operating frequency of $f_0 = 11.994$ GHz. The group velocity obtained is $v_g = 0.066c$, where $c$ is speed of light. An unloaded quality factor of $Q_0 = 2829$ and a shunt impedance of $R_{\text{shunt}} = 26.5$ MΩ/m are also derived for such a DLA structure. Both RF parameters are much less advantageous when compared to existing metallic CLIC [30-32] structures, agreeing with the prediction of a large $H_p/E_a = 9.32$ mA/V. The length of the DLA structure is chosen as 100 mm for the following simulations and mechanical assembly.

# III. Design of a dielectric matching section

Simulation studies [36] have shown that a good match can be achieved for the impedance of the $TM_{01}$ mode between the circular waveguide and the dielectric-loaded waveguide by using a vacuum-waveguide quarter-wavelength matching section. However, there is an issue with very high electromagnetic fields located in the vacuum-waveguide matching section due to strong resonance within it. In addition, it has a very narrow bandwidth. As a solution, a dielectric disk is therefore added into the vacuum matching section. In this section, a dielectric-based quarter-wavelength matching section to efficiently couple RF power from a circular waveguide into the DLA structure will be proposed and studied in detail.

## A. Rectangular angle

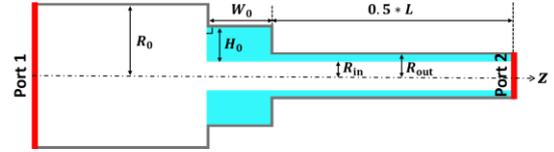

Figure 5. Longitudinal cross section of a circular waveguide, a dielectric matching section and a DLA structure.

The DLA structure has the geometry listed in Table 1. Figure 5 shows the dielectric-based matching section which consists of a dielectric cylindrical tube with a rectangular angle. The dielectric material is exactly the same as the DLA structure. This matching section has a width $W_0$ and a height $H_0$ for a fixed $R_{\text{in}} = 3.0$ mm. There are two waveguide ports (Port 1 and Port 2, see Fig. 5) defined for the calculation of S-parameters in HFSS simulations. The desired reflection coefficient $S_{11}$ and transmission coefficient $S_{21}$ can be realized by tuning the value of $W_0$ and $H_0$.

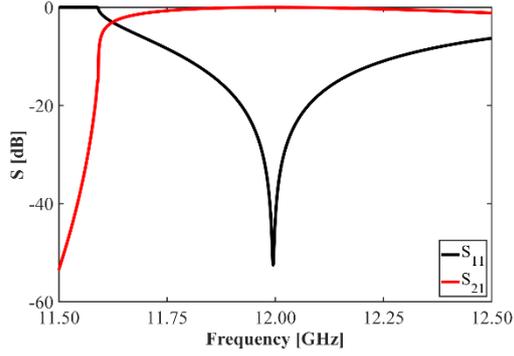

Figure 6. Simulated $S_{11}$ and $S_{21}$ as a function of frequency for an optimum geometry shown in Fig. 5.

After optimization, values of $S_{11} = -51.5$ dB and $S_{21} = -0.03$ dB are achieved at the operating frequency of 11.994 GHz for a dielectric matching section with $W_0 = 1.777$ mm and $W_0 = 3.1$ mm, as shown in Fig. 6. This dielectric matching section has a width much smaller than that of previously-reported tapered matching sections [22-23], since it bases on different matching mechanism: a quarter-wavelength transformer. The $S_{11} = -51.5$ dB indicates that the reflected RF power is negligibly small. Using $S_{21} = -0.03$ dB, the coupling coefficient is calculated to be 99.3%. This means that almost 100% of RF power is efficiently coupled from the circular waveguide into the DLA structure by using a compact dielectric matching section such as this, with a rectangular angle. The $S_{21}$ also has a broad 3 dB bandwidth of more than 1.0 GHz, which allows greater tolerance to potential fabrication errors. The following design based on these S-parameters shows great promise.

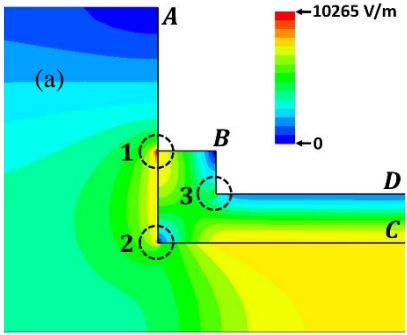

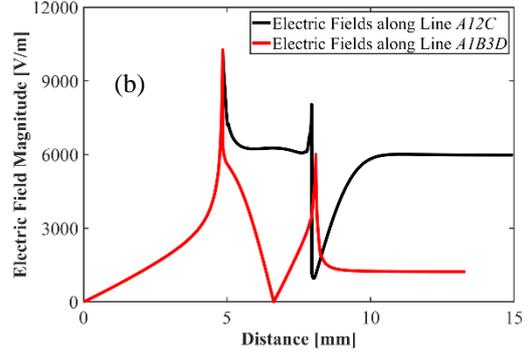

Figure 7. (a) Electric field distribution for the geometry shown in Fig. 5. (b) Electric field magnitude along Line *A12C* and Line *A1B3D*, respectively. Here Line *A12C* and Line *A1B3D* denote a section of lines connected by the points *A*, *1*, *2*, *C*, *B*, *3*, *D*, as shown in (a), where the distance of point *A* is taken as 0 mm.

After simulating S-parameters, we must now investigate the RF field distribution for this compact dielectric-based matching section. Only electric fields are analysed here because it is they which limit the achievable accelerating gradient for the DLA structures. Figure 7 (a) shows the electric field distribution for the dielectric matching section connected with a circular waveguide and a DLA structure, at an input power of 1.0 W. It is found that very strong fields are located at three corners denoted by the numbers 1, 2, and 3, as shown in Fig. 7 (a). Figure 7 (b) shows the electric field magnitude along Line *A12C* and Line *A1B3D* (see Fig. 7 (a)), respectively. A total of three peaks are observed in these black and red curves, corresponding to the strong fields at Corner 1, Corner 2 and Corner 3. Solutions are required to reduce the strong fields at these three corners.

**B. Tilt angle**

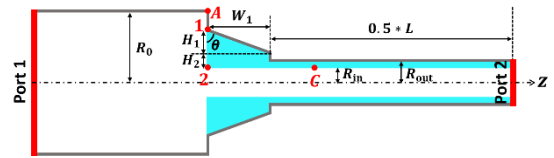

Figure 8. Longitudinal cross section of a circular waveguide, a tilt dielectric matching section and a DLA structure. Here Line *A12C* has the same definition as in Fig. 7(a).

In order to reduce the peak fields at Corner 1, the dielectric matching section with a rectangular angle can be changed to a tilt dielectric matching section, as shown in Fig. 8. It has four geometrical parameters: a width $W_1$, an upper height $H_1$, a lower height $H_2$, and a tilt angle $\theta$. When two arbitrary variables are selected from $W_1$, $H_1$ and $\theta$, the remaining one is also fixed. Together with $H_2$, the desired S-parameters ($S_{11}$ and $S_{21}$) can be obtained by tuning a combination of the three variables.

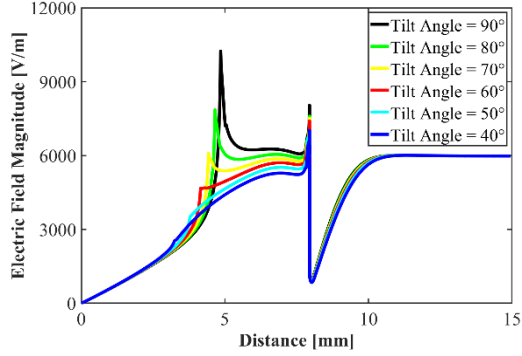

Figure 9. Electric field magnitude along Line $A12C$ (see Fig. 8) for the dielectric matching section, for different tilt angles.

Figure 9 shows the calculated electric field magnitude along Line $A12C$ (see Fig. 8) for the dielectric matching section for different tilt angles $\theta$, at an input power of 1.0 W. When tilt angle $\theta = 90^0$, this is the rectangular angle case, which has been described in subsection A. There are two obvious peaks for each simulated curve. The left peak indicates the strong fields at Corner 1 while the right peak denotes the strong fields at Corner 2. As expected for this kind of triple-point, it can clearly be seen that the left peak gradually drops with a smaller tilt angle, while the right peak still exists. For a tilt angle $\theta = 60^0$, the peak fields at Corner 1 are reduced below those of the DLA structure. Such a tilt angle can be also realised without fabrication difficulties. A tilt angle $\theta = 60^0$ is therefore chosen for our dielectric matching section.

## C. Corner rounding

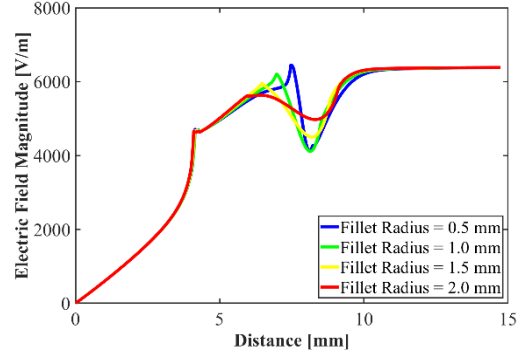

Figure 10. Electric field magnitude along Line $A1C$ (see Fig. 11 (a)) for different fillet radii at Corner 2.

As seen in Fig. 9, the right peak corresponding to the strong fields located at Corner 2 still exists for a tilt dielectric matching section. In order to reduce the peak fields at Corner 2, we round this corner by a fillet radius $R_f$. Figure 10 shows the influence of different fillet radii on the electric field magnitude at Corner 2, at an input power of 1.0 W. The peak fields at Corner 2 gradually become weaker with a larger fillet radius. Therefore, a fillet radius of $R_f = 2.0$ mm is chosen for Corner 2, resulting in electric fields along Line $A1C$ (see Fig. 11) much lower than those of the DLA structure.

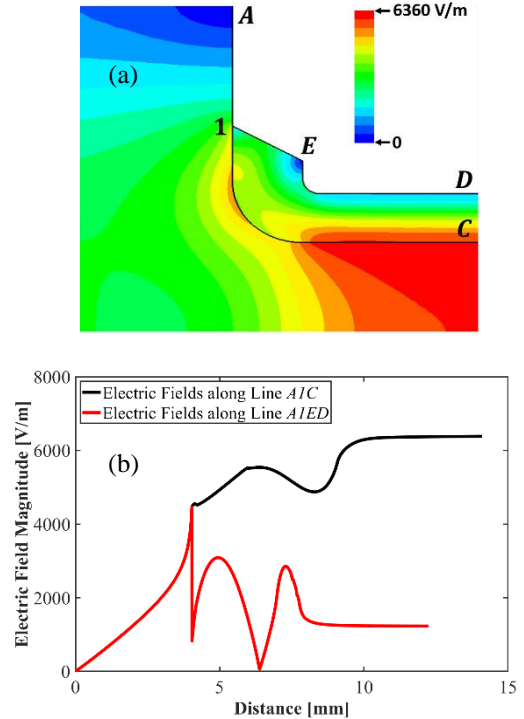

Figure 11. (a) Electric field distribution for the optimum dielectric matching section. (b) Electric field magnitude

along Line $A1C$ and Line $A1ED$. Here Line $A1C$ and Line $A1ED$ denote a section of lines and arcs connected by the points $A$, $1$, $C$, $D$, and $E$, as shown in (a), where the distance of point $A$ is taken as 0 mm.

A similar rounding method is also applied for Corner 3 with a fillet radius of $R_m = 0.5$ mm. After such a rounding, we find the electric field distribution shown in Fig. 11 (a) in the matching section at an input power of 1.0 W. The electric fields at Corner 3 are much weaker than those of the DLA structure, as shown in red curve of Fig. 11 (b). Therefore, our dielectric matching section with a tilt angle of $\theta = 60^0$, and rounded Corners 2 and 3 with fillet radii of $R_f = 2.0$ mm and $R_m = 0.5$ mm, respectively, has electric fields much lower than those of the DLA structure. In this case, such a compact dielectric matching section has the potential to survive in the high-power environment, as the maximum electric fields are located in the DLA structure, and does not limit the high gradient performance of the DLA structure itself.

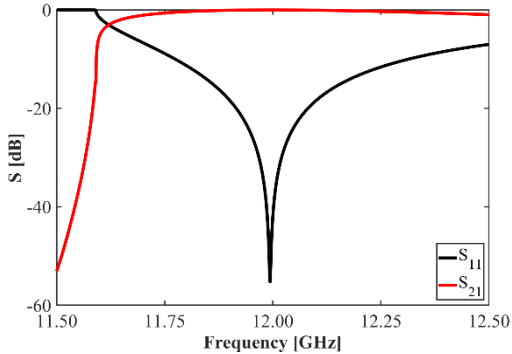

Figure 12. Simulated $S_{11}$ and $S_{21}$ as a function of frequency for the optimum dielectric matching section.

For a dielectric matching section with a tilt angle of $\theta = 60^0$, and rounded Corners 2 and 3 with fillet radii of $R_f = 2.0$ mm and $R_m = 0.5$ mm, respectively, an optimum geometry can be found by adjusting width $W_1$ and lower height $H_2$ together, to provide a good matching between the circular waveguide and the DLA structure. Figure 12 shows the calculated $S_{11} = -55$ dB and $S_{21} = -0.03$ dB for an optimum dielectric matching section with a width of $W_1 = 2.031$ mm and a lower height of $H_2 = 2.743$ mm, at the operating frequency of 11.994 GHz. Both $S_{11}$ and $S_{21}$ are almost exactly the same as for the matching section with a rectangular angle. It has a coupling coefficient of 99.3% and $S_{21}$ also has a broad 3 dB bandwidth of more than 1.0 GHz.

## IV. Considerations for realistic fabrication studies

In this section, we take realistic fabrication requirements into account for our design. These fabrication requirements, including chamfers on sharp dielectric corners and the existence of microscale vacuum gaps caused by metallic clamping, are studied in simulations.

### A. Chamfer the sharp corners

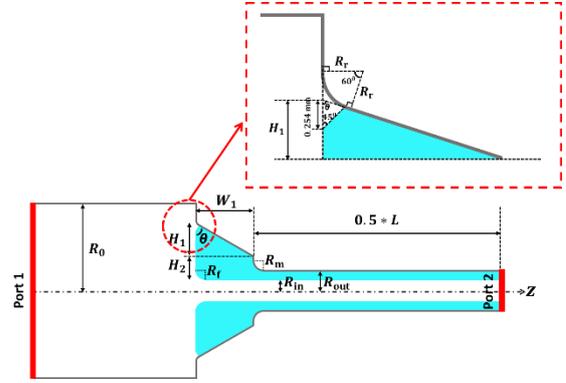

Figure 13. Longitudinal cross section of a circular waveguide, an optimum dielectric matching section with a chamfered dielectric corner, and a DLA structure.

In the realistic fabrication of the dielectric matching section, a sharp corner is easily broken. In order to prevent such a break, a $45^0$ chamfer with a length of 0.254 mm is added for the sharp dielectric corner, as shown in Fig. 13. The shape of the outer metal is also changed by rounding with a fillet radius of $R_r = 0.322$ mm, in order to prevent field enhancement near that area. The width $W_1$, the upper height $H_1$, and the lower height $H_2$ of the dielectric matching section are thereby adjusted for the best matching performance and RF field distribution.

After chamfering the sharp dielectric corner, it is found in simulations that the width $W_1$ is changed from 2.031 mm to 2.035 mm while the lower height $H_2$ is adjusted from 2.743 mm to 2.74 mm. A dielectric matching section with realistic geometry is therefore obtained as follows: $W_1 = 2.035$ mm, $H_2 = 2.74$ mm, $\theta = 60^0$, $R_f = 2.0$ mm, and $R_m = 0.5$ mm. Both $S_{11}$ and $S_{21}$ have the same properties as the optimum matching section in terms of a 3 dB bandwidth. This means that the chamfer on the sharp dielectric corner does not have an influence on the coupling coefficient after optimizations.

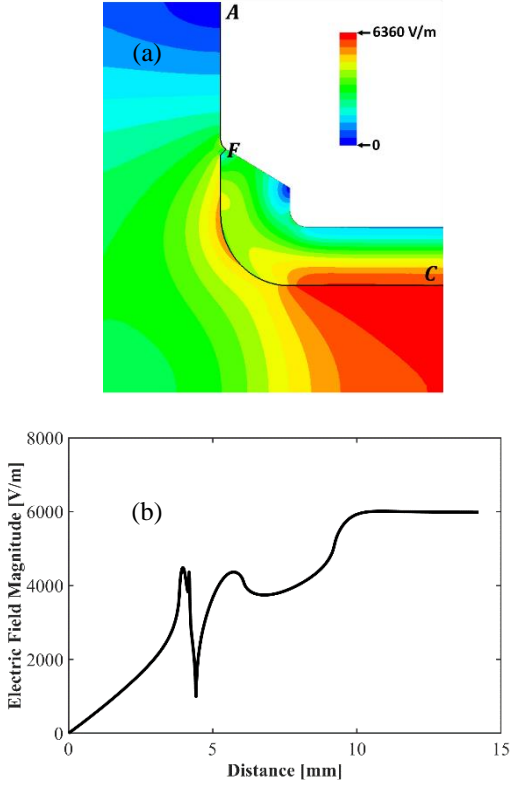

Figure 14. (a) Electric field distribution for the optimum matching section with a chamfered dielectric corner. (b) Electric field magnitude along Line *AFC* which denotes a section of lines and arcs connected by the points *A*, *F*, and *C*, as shown in (a), where the distance of point *A* is taken as 0 mm.

Figure 14 (a) shows the calculated electric field distribution for the dielectric matching section with a realistic geometry at an input power of 1.0 W. Figure 14 (b) indicates that the electric fields located in a realistic matching section are much lower than those of the DLA structure. It can be clearly seen that the strongest fields, with an amplitude of 6360 V/m, are located in the DLA structure, which is exactly the same as in the optimum matching section (see Fig. 11 (a)). This is the ideal case for future high-power tests on our DLA structure.

## B. A vacuum microgap

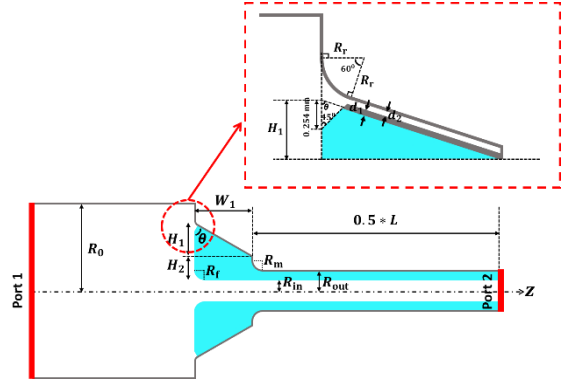

Figure 15. Longitudinal cross section of a circular waveguide, a realistic dielectric matching section with a vacuum microgap, and a DLA structure.

It was found [14] that any microgap caused by a dielectric joint resulted in RF breakdown, due to strong field enhancement. In our realistic fabrication the entire dielectric tube, including the matching section and the DLA structure, is therefore sintered as a single piece. A thin metallic layer of 0.0508 mm is first coated onto the surface of the whole dielectric tube. The coated dielectric tube is then inserted into the outer copper jacket. However, there is still a microscale vacuum gap caused by metallic clamping between the thin metallic coating and the outer thick copper jacket, as shown in Fig. 15. It is therefore particularly import to study the dependence of the S-parameters and electric fields on the microgap $d_2$.

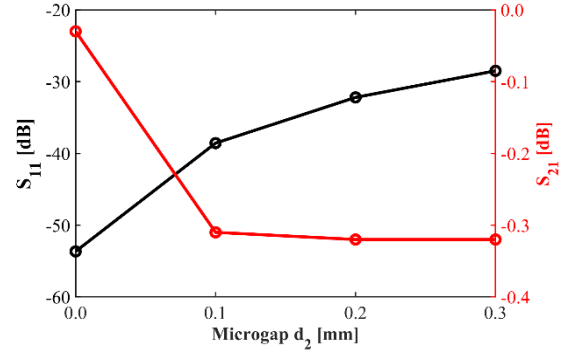

Figure 16. Simulated $S_{11}$ and $S_{21}$ as a function of vacuum microgap $d_2$.

Figure 16 shows how varying the vacuum microgap $d_2$ influences $S_{11}$ and $S_{21}$. With a larger vacuum microgap, $S_{11}$ increases while $S_{21}$ decreases, resulting in worse matching. For a vacuum microgap of $d_2 = 0.2$ mm, $S_{11} = -32$ dB and $S_{21} = -0.32$ dB are obtained, which is still acceptable for our design. However, $S_{11}$ is increased to -28.5 dB and $S_{21}$ remains unchanged, when the vacuum microgap becomes 0.3 mm.

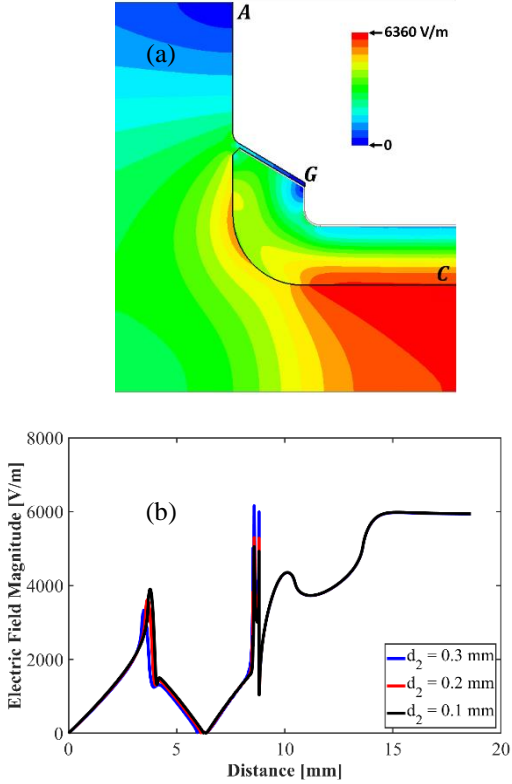

Figure 17. (a) Electric field distribution for the realistic dielectric matching section with a vacuum microgap, where the thin metallic coating is denoted by the white lines. (b) Electric fields magnitude along Line *AGC* for different vacuum microgaps $d_2$. Here Line *AGC* denotes a section of lines and arcs connected by the points *A*, *G*, and *C*, as shown in (a), where the distance of point *A* is taken as 0 mm.

Figure 17 (a) shows the calculated electric field distribution for the realistic matching section with a vacuum microgap, at an input power of 1.0 W. Figure 17 (b) gives the calculated electric field magnitude along Line *AGC* for different vacuum microgaps. There are two peaks in each curve, indicating the relatively strong fields near the chamfered corner and rounding Corner 2, respectively. For a vacuum microgap of 0.3 mm, the peak fields are higher than those of the DLA structure, which may cause arcing in a high-power test. The dielectric matching section is therefore allowed to have a maximum vacuum microgap of 0.2 mm, in which RF fields are still lower than those of DLA structure, $S_{11}$ is better than -30 dB, and the coupling coefficient is 93%. This value is used to guide the fabrication tolerances of the copper jacket and the metallic coating of the dielectric tube.

## V. Tolerance studies

Through previous studies, realistic geometrical parameters for the dielectric matching section and the DLA structure are obtained as follows: $\varepsilon_r = 16.66$, $W_1 = 2.035$ mm, $H_2 = 2.74$ mm, $\theta = 60^0$, $R_f = 2.0$ mm, $R_m = 0.5$ mm, $R_{out} = 4.6388$ mm, $R_{in} = 3.0$ mm, and $L = 100$ mm. Using these geometrical parameters, $S_{11} = -54$ dB and $S_{21} = -0.03$ dB are achieved at the operating frequency of 11.994 GHz. The length of the DLA structure does not have any effect on the S-parameters and RF-field performance, so it is ruled out for tolerance studies in this section. The tolerances of key geometrical parameters (see Table 2) are discussed in detail.

Table 2. The tolerances of geometrical parameters for the dielectric matching section and DLA structure.

| $f_0 = 11.994$ GHz | $S_{11} \leq -30$ dB | $S_{11} \leq -25$ dB | $S_{11} \leq -20$ dB |
|---|---|---|---|
| $\varepsilon_r = 16.66$ | [-0.079, +0.081] | [-0.139, +0.148] | [-0.24, +0.27] |
| $W_1 = 2.035$ [mm] | [-0.007, +0.007] | [-0.012, +0.012] | [-0.022, +0.022] |
| $H_2 = 2.74$ [mm] | [-0.015, +0.017] | [-0.027, +0.030] | [-0.051, +0.054] |
| $\theta = 60^0$ | [-2.5$^0$, +2.0$^0$] | [-4.3$^0$, +3.7$^0$] | [-7.3$^0$, +7.0$^0$] |
| $R_f = 2.0$ [mm] | [-0.042, +0.040] | [-0.076, +0.068] | [-0.140, +0.120] |
| $R_m = 0.5$ [mm] | [-0.061, +0.049] | [-0.118, +0.090] | [-0.245, +0.151] |
| $R_{out} = 4.6388$ [mm] | [-0.0076, +0.0065] | [-0.0123, +0.0127] | [-0.020, +0.025] |
| $R_{in} = 3.0$ [mm] | [-0.006, +0.007] | [-0.012, +0.012] | [-0.024, +0.020] |

As we know, $S_{21}$ for the realistic matching section has a large 3 dB bandwidth of over 1 GHz, so it is not sensitive to changes in the geometrical parameters. The tolerances are studied by calculating the dependence of $S_{11}$ on the geometrical parameters. By adjusting a certain geometrical parameter from *x* to *x* ± *dx*, $S_{11}$ is calculated and compared with the setting requirements of -30 dB, -25 dB, and -20 dB. As shown in Table 2, $S_{11}$ is very sensitive to $W_1$, $R_{out}$, and $R_{in}$ and less sensitive to $\varepsilon_r$, $H_2$, $\theta$, $R_f$, and $R_m$. The dielectric fabrication accuracy should be better than ± 0.02 mm in order to realize a $S_{11} \leq -20$ dB, which is still acceptable for efficient coupling.

# VI. Full-assembly structure

In this section, a full-assembly structure is obtained by adding the DLA structure connected together with two matching sections, circular waveguides with the choke geometry, and the $TE_{10}$-$TM_{01}$ mode converters. The RF performance of such a full assembly structure is described in detail.

## A. A choke geometry

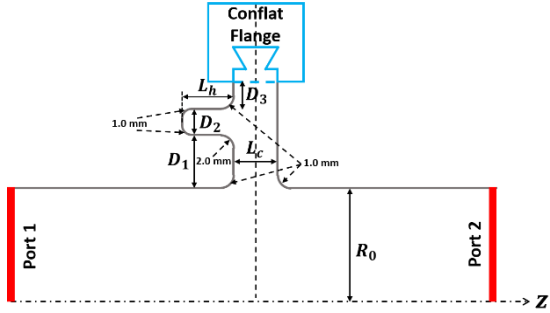

Figure 18. Longitudinal cross section of a circular waveguide with a choke geometry.

In order to remove the contact issue for assembling two parts together, a choke geometry is added, as shown in Fig. 18. In this case, the disks have no bonding joints in high RF fields. There are five geometrical parameters ($D_1$, $D_2$, $D_3$, $L_c$, and $L_h$) for a choke geometry. Through proper design, this choke can be used to reflect the fundamental $TM_{01}$ mode back to the circular waveguide so that it will not affect the RF fields travelling in the circular waveguide. A conflat flange, located outside, is used to assemble both of the circular vacuum waveguides tightly together.

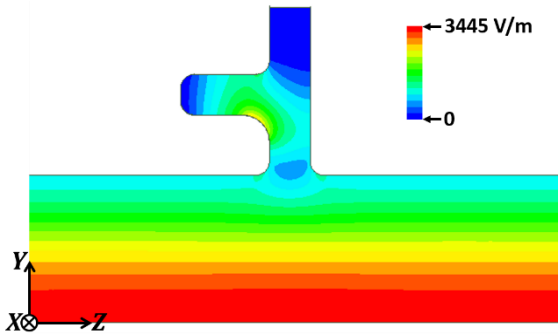

Figure 19. Electric field distribution for the accelerating mode $TM_{01}$ in a circular waveguide with a choke geometry.

After simulations, the optimum is obtained with a choke having dimensions $D_1 = 4.5$ mm, $D_2 = 3.0$ mm, $D_2 \geq 5.0$ mm, $L_c = 3.0$ mm, $L_h = 6.535$ mm. Fillet radii are also added to round all of the corners, as seen in Fig. 19. Figure 19 shows the electric field distribution for the fundamental $TM_{01}$ mode in a circular waveguide with this optimum choke, at an input power of 1.0 W. It can be clearly seen that the electric fields in such an optimum choke are much weaker than those in the circular waveguide. This indicates the choke should not affect the high-power performance of the circular waveguide for operation in $TM_{01}$ mode.

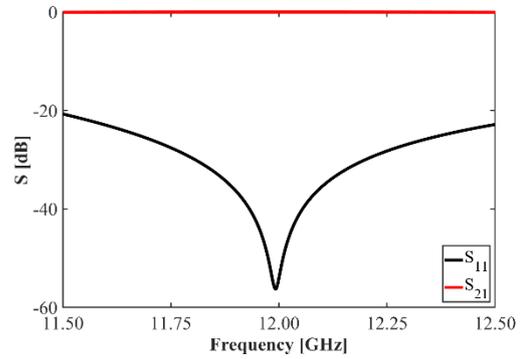

Figure 20. Simulated $S_{11}$ and $S_{21}$ as a function of frequency for a choke geometry connected to a circular waveguide, as shown in Fig. 18.

Figure 20 gives calculated values of $S_{11} = -56$ dB and $S_{21} = -0.01$ dB for a circular waveguide with a choke geometry, at an operating frequency of 11.994 GHz. This means that 99.8% of RF power is transmitted from Port 1 to Port 2 (see Fig. 18) and 0.2% of RF power is dissipated on the metallic surface and choke structure. The optimum choke has a negligible effect on the RF power transmitted in the circular waveguide. In addition, the transmission coefficient $S_{21}$ for this choke has a very broad 3 dB bandwidth of more than 3 GHz. This allows flexibility in the fabrication and mechanical assembly requirements for a choke geometry.

## B. Full-assembly structure

A $TE_{10}$-$TM_{01}$ mode converter at a frequency of 11.994 GHz has been studied at CERN [24]. It is used to convert the $TE_{10}$ mode from a rectangular waveguide to the $TM_{01}$ mode in a circular waveguide. By using this existing mode converter, together with the choke geometry, the realistic matching section and the DLA structure, a full-assembly structure as shown in Fig. 21 is obtained. We simulate the whole structure by analysing the electric field distribution and S-parameters from Port 1 to Port 2. RF power loss, on both the metallic surface and in dielectrics and accelerating fields in the vacuum channel, are also studied in detail.

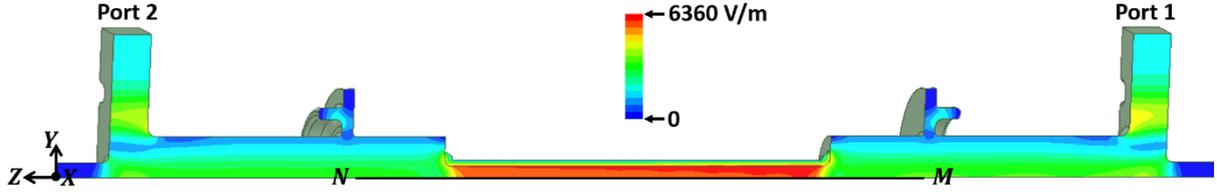

Figure 21. Electric field distribution for the full-assembly structure, where line *MN* is located on the centre along *z*-axis.

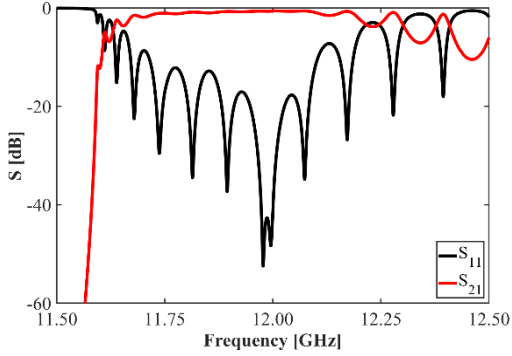

Figure 22. Simulated $S_{11}$ and $S_{21}$ as a function of frequency for the full-assembly structure shown in Fig. 21.

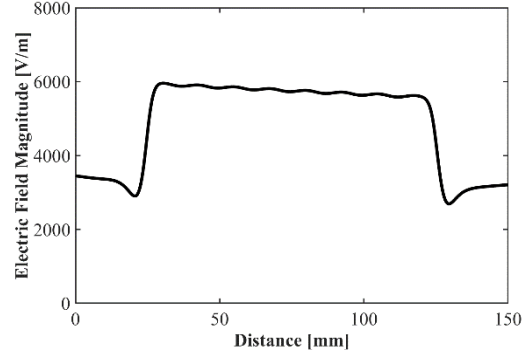

Figure 23. Electric field magnitude along the line *MN* shown in Fig. 21. The distance of point *M* is taken as 0 mm.

Figure 22 gives calculated values of $S_{11} = -47.5$ dB and $S_{21} = -0.67$ dB for the full-assembly structure at the operating frequency of 11.994 GHz. Using the power density on the metallic surface and in the dielectric area for an input power of 1.0 W, the calculated RF power loss on the metallic surface is $P_{\text{loss\_surface}} = 0.130$ W and the RF power loss obtained in dielectrics is $P_{\text{loss\_dielectric}} = 0.012$ W. So the total RF power loss is $P_{\text{total\_loss}} = 0.142$ W. The output RF power at Port 2 is $P_{\text{out}} = 0.858$ W. We thus achieve a transmission coefficient $S'_{21} = 10 \log(P_{\text{out}}/P_{\text{in}}) = -0.67$ dB, which agrees well with the simulated $S_{21}$ shown in Fig. 22. At the maximum peak power of 40 MW from XBOX [37-38] with a pulse width of 1.5 μs and a repetition rate of 50 Hz an average input power of 3.0 kW will be generated. The power loss on the metallic surface is then 390 W and the power loss in dielectrics is 36 W. A water cooling system is thus required for the high-power test on the full-assembly structure.

Figure 23 shows the electric field magnitude along a line *MN* (see Fig. 21). The electric fields are gradually becoming weaker, due to RF power loss in the dielectric and on metallic surfaces, as the RF fields propagate from point *M* to point *N*. The average accelerating gradient is calculated to be 5773 V/m at an input power of 1.0 W. For a power of 40 MW from XBOX, an average accelerating gradient of 36.5 MV/m can be achieved for our DLA structure.

Figure 24 presents the full-assembly mechanical design for the whole structure. The grey area denotes the outer copper jacket, connected with openings to avoid air trapping when pumping. Two conflat flanges are used to connect the centre part with the end parts, which are $TE_{10}$-$TM_{01}$ mode converters.

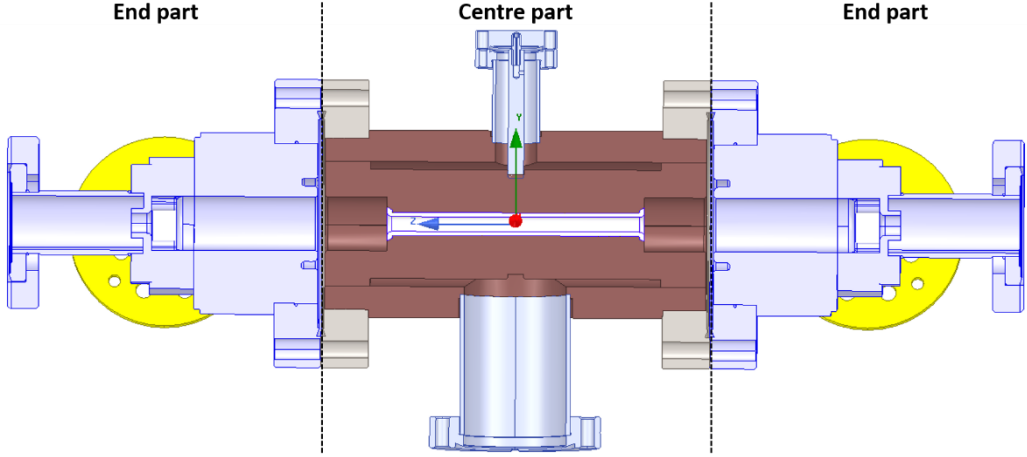

Figure 24. Full-assembly mechanical design for the whole structure, including a centre part and two end parts

## VII. Conclusion

In this paper, a compact, low-field, broadband dielectric matching section has been proposed and studied to efficiently couple the RF power from a circular waveguide to an X-band DLA structure. Through simulation studies, an optimum dielectric matching section with a tilt angle of $\theta = 60^0$, a width of $W_1 = 2.031$ mm and a lower height of $H_2 = 2.743$ mm is obtained, achieving a reflection coefficient of $S_{11} = -55$ dB and a transmission coefficient of $S_{21} = -0.03$ dB with a very broad 3 dB width of more than 1 GHz. In order to prevent a sharp dielectric corner break, a $45^0$ chamfer with a length of 0.254 mm is added. It is also found that the maximum allowable size for any microscale vacuum gap caused by metallic clamping between a thin metallic coating and the outer thick copper jacket is 0.2 mm. In this case, the RF fields in the matching section are lower than those of the DLA structure, and the reflection coefficient is better than -30 dB. Tolerance studies show that the dielectric fabrication accuracy should be better than ± 0.02 mm in order to realize a $S_{11} \leq -20$ dB. Finally, a full-assembly structure, including the DLA structure connected with two matching sections, circular waveguides with the choke geometry, and the $TE_{10}$-$TM_{01}$ mode converters, is analysed in detail. $S_{11} = -47.5$ dB and $S_{21} = -0.67$ dB are obtained for this full-assembly structure. For a peak power of 40 MW with a pulse width of 1.5 μs and a repetition rate of 50 Hz, the power loss on the metallic surface is 390 W and the power loss in dielectrics is 36 W. A water cooling system is thus required for its high-power test. An average accelerating gradient of 36.5 MV/m can be also expected for our DLA structure.

The centre part, including the whole dielectric tube with the outer copper jacket, is under construction at Euclid Techlabs, while the fabrication of mode converters is underway at CERN. After fabrication and mechanical assembly, the whole structure will be tested and benchmarked with results from simulations, to improve these designs.

## Acknowledgments

The authors would like to thank Dr. Walter Wuensch for the useful comments, the team of Argonne Wakefield Accelerator facility (Dr. Manoel Conde, Dr. John Power, and Dr. Jiahang Shao, etc) for the fruitful discussions, Dr. Joel Sauza Bedolla for the mechanical design of the mode converters, and Dr. Mark Ibison for his careful reading of the manuscript.

[1] R. B. R.-Shersby-Harvie, A proposed new form of dielectric-loaded wave-guide for linear electron accelerators, Nature 162, 890 (1948).

[2] G. T. Flesher and G. I. Cohn, Dielectric loading for waveguide linear accelerators, Trans. Am. Inst. Electr. Eng. 70, 887 (1951).

[3] R. B. R.-Shersby-Harvie, L. Mullett, W. Walkinshaw, J. Bell and B. Loach, A theoretical and experimental investigation of anisotropic-dielectric-loaded linear electron accelerators, Proc. IEE B 104, 273 (1957).

[4] L. Xiao, W. Gai, and X. Sun, Field analysis of a dielectric-loaded rectangular waveguide

accelerating structure, in *Proceedings of the 2001 Particle Accelerator Conference*, Chicago, IL, USA, (IEEE, Piscataway, NJ, 2001), pp. 3963-3965.

[5] C. Jing, W. M. Liu, W. Gai, J. G. Power, T. Wong, Mode analysis of a multilayered dielectric-loaded accelerating structure, Nucl. Instrum. Methods Phys. Res., Sect. A 539, 445 (2005).

[6] G. B. Walker and E. L. Lewis, Vacuum Breakdown in Dielectric-loaded Wave-guides, Nature 181, 38 (1958).

[7] W. Gai, P. Schoessow, B. Cole, R. Konecny, J. Norem, J. Rosenzweig, and J. Simpson, Experimental Demonstration of Wake-Field Effects in Dielectric Structures, Phys. Rev. Lett. 61, 2756 (1988).

[8] J. G. Power, W. Gai, S. H. Gold, A. K. Kinkead, R. Konecny, C. Jing, W. Liu, and Z. Yusof, Observation of Multipactor in an Alumina-Based Dielectric-Loaded Accelerating Structure, Phys. Rev. Lett. 92, 164801 (2004).

[9] O. Sinitsyn, G. Nusinovich, and T. Antonsen, Studies of multipactor in dielectric - loaded accelerator structures: comparison of simulation results with experimental data, AIP Conf. Proc. 1299, pp. 302-306 (2010).

[10] D. Satoh, M. Yoshida, and N. Hayashizaki, Fabrication and cold test of dielectric assist accelerating structure, Phys. Rev. Accel. Beams 20, 091302 (2017).

[11] W. Gai and C. Ho, Modeling of the transverse mode suppressor for dielectric wake‐field accelerator, J. Appl. Phys. 70, 3955 (1991).

[12] E. Chojnacki, W. Gai, C. Ho, R. Konecny, S. Mtingwa, J. Norem, M. Rosing, P. Schoessow, and J. Simpson, Measurement of deflection - mode damping in an accelerating structure, J. Appl. Phys. 69, 6257 (1991).

[13] W. K. H. Panofsky and M. Bander, Asymptotic Theory of Beam Break‐Up in Linear Accelerators, Rev. Sci. Instrum. 39, 206 (1968).

[14] C. Jing, W. Gai, J. G. Power, R. Konecny, S. H. Gold, W. Liu, and A. K. Kinkead, High-power RF tests on X-band external powered dielectric-loaded accelerating structures, IEEE Trans. Plasma Sci. 33, 1155 (2005).

[15] C. Jing, W. Gai, J. G. Power, R. Konecny, W. Liu, S. H. Gold, A. K. Kinkead, S. G. Tantawi, V. Dolgashev, and A. Kanareykin, Progress toward externally powered X-band dielectric-loaded accelerating structures, IEEE Trans. Plasma Sci. 38, 1354 (2010).

[16] C. Chang, J. Verboncoeur, S. Tantawi and C. Jing, The effects of magnetic field on single-surfaceresonant multipactor, J. Appl. Phys. 110, 063304 (2011).

[17] C. Jing, C. Chang, S. H. Gold, R. Konecny, S. Antipov, P. Schoessow, A. Kanareykin, and W. Gai, Observation of multipactor suppression in a dielectric-loaded accelerating structure using an applied axial magnetic field, Appl. Phys. Lett. 103, 213503 (2013).

[18] C. Jing, S. H. Gold, R. Fischer and W. Gai, Complete multipactor suppression in an X-band dielectric-loaded accelerating structure, Appl. Phys. Lett. 108, 193501 (2016).

[19] M. C. Thompson, H. Badakov, A. M. Cook, J. B. Rosenzweig, R. Tikhoplav, G. Travish, I. Blumenfeld, M. J. Hogan, R. Ischebeck, N. Kirby, R. Siemann, D. Walz, P. Muggli, A. Scott, and R. B. Yoder, Breakdown Limits on Gigavolt-per-Meter Electron-Beam-Driven Wakefields in Dielectric Structures, Phys. Rev. Lett. 100, 214801 (2008).

[20] P. Zou, W. Gai, R. Konecny, X. Sun, T. Wong and A. Kanareykin, Construction and testing of an 11.4 GHz dielectric structure based traveling wave accelerator, Rev. Sci. Instrum. 71, 2301 (2000).

[21] J. G. Power, W. Gai, C. Jing, R. Konecny, S. H. Gold, A. K. Kinkead, High power testing of ANL X-band dielectric-loaded accelerating structures, AIP Conf. Proc. 647, pp. 156-164 (2002).

[22] W. Liu and W. Gai, Design of dielectric accelerator using TE-TM mode converter, AIP Conf. Proc. 647, pp. 469-475 (2002).

[23] W. Liu, C. Jing, W. Gai, R. Konecny, and J. G. Power, New RF design for 11.4 GHz dielectric loaded accelerator, in *Proceedings of the 2003 Particle Accelerator Conference, Portland, Oregon, USA* (IEEE, Piscataway, NJ, 2003), pp. 1810–1812.

[24] I. Syratchev, Mode launcher as an alternative approach to the cavity-based RF coupler of periodic structures, CERN-OPEN-2002-005 (2002).

[25] A. Kanareykin, New Advanced Dielectric Materials for Accelerator Applications, AIP Conf. Proc. 1299, pp. 286-291 (2010).

[26] S. H. Gold, C. Jing, A. Kanareykin, W. Gai, R. Konecny, J. G. Power, and A. K. Kinkead, Development of X-band dielectric - loaded

accelerating structures, AIP Conf. Proc. 1299, pp. 292-296 (2010).

[27] HFSS, www.ansys.com.

[28] P. B. Wilson, RF-Driver Linear Colliders, in *Proceedings of the 1987 Particle Accelerator Conference, Washington, D.C., USA* (IEEE, Piscataway, NJ, 1987), pp. 53–58.

[29] J. W. Wang *et al.*, Accelerator Structure R&D for Linear Colliders, in *Proceedings of the 1999 Particle Accelerator Conference, New York, USA* (IEEE, Piscataway, NJ, 1999), pp. 3423–3425.

[30] A. Grudiev, W. Wuensch, Design of the CLIC main linac accelerating structure for CLIC Conceptual Design Report, in *Proceedings of the 25th International Linear Accelerator Conference, LINAC2010, Tsukuba, Japan* (KEK, Tsukuba, Japan, 2010).

[31] H. Zha and A. Grudiev, Design and optimization of Compact Linear Collider main linac accelerating structure, Phys. Rev. Accel. Beams 19, 111003 (2016).

[32] H. Zha and A. Grudiev, Design of the Compact Linear Collider main linac accelerating structure made from two halves, Phys. Rev. Accel. Beams 20, 042001 (2017).

[33] I. Wilson, Surface heating of the CLIC main linac structure, CLIC-Note-52 (1987)

[34] D. P. Pritzkau, "RF pulsed heating," Ph.D. thesis, Stanford University, 2001.

[35] D. P. Pritzkau and R. H. Siemann, Experimental study of rf pulsed heating on oxygen free electronic copper, Phys. Rev. ST Accel. Beams 5, 112002 (2002).

[36] Y. Wei, A compact, low-field, broadband mode launcher for X-band DLA structures, Presentation of AWA Needs and Opportunities Workshop in Argonne National Laboratory, USA ( 2019 ), https://indico.fnal.gov/event/20265/contribution/25.

[37] N. Catalan-Lasheras, A. Degiovanni, S. Doebert, W. Farabolini, J. Kovermann, G. McMonagle, S. Rey, I. Syratchev, L. Timeo, W. Wuensch, B. Woolley, J. Tagg, Experience operating an X-band high-power test stand at CERN, in Proceedings of the 5th International Particle Accelerator Conference, IPAC2014, Dresden, Germany (JACoW, Geneva, 2014), pp. 2288–2290.

[38] A. V. Edwards, N. Catalan Lasheras, S. Gonzalez Anton, G. Mcmonagle, S. Pitman, V. del Pozo Romano, B. Woolley, A. Dexter, Connection of 12 GHz high power RF from the XBOX1 high gradient test stand to the CLEAR electron linac, in Proceedings of the 10th International Particle Accelerator Conference, IPAC2019, Melbourne, Australia (JACoW, Geneva, 2019), pp. 2960-2963.